# A novel dynamic scheme to reconstruct sub-50 atto-second pulses

Renzhi Shao[1,2], Bin Li[1,2,3*]

[1]*Shanghai Institute of Applied Physics, Chinese Academy of Sciences, Shanghai 201204, China*
[2]*University of Chinese Academy of Sciences, Beijing 100049，China*
[3]*ShanghaiTech University, School of Physics Science and Technology, Shanghai 201210，China*
[*]*Corresponding Author: libin1995@sinap.ac.cn*

**Abstract**：The attosecond streaking technique is normally utilized to characterize the attosecond pulses in extreme ultraviolet spectrum, while the pulse simulation and reconstruction schemes turn out to be unreliable for extremely short sub-100as pulses. Based on systematical analysis and subtle investigation, we developed a dynamic streaking model for ultrashort photoelectron pulses and demonstrated concretely how the temporal phase of attosecond pulses would influence the streaked spectrogram. Upon that, a novel numeric scheme is established to retrieve the pulse-lengths and chirps. Different with the traditional FROG-CRAB algorithm utilizing the central energy approximation, our approach concerns the actual time-energy distribution of an electron wave-packet within sub-100as time-scale, capable for retrieving the temporal profiles of the ultrashort pulse more accurately.

1. Introduction

The cutting edge atto-second laser science and technology lead to the breakthrough to sub-100as pulse generation[1, 2]. Accurate characterization of the attosecond pulse is particularly important for the atto-second sciences and applications[3, 4]. The two-color experiment is usually applied to measure and reconstruct the attosecond pulse, which typically in either extreme ultraviolet (XUV) or soft X-ray spectral range, ionized an atom in the presence of a synchronized dressing laser field in infrared (IR). Where the pulse length of ionized photoelectrons could be regarded as mapping and identical to that of the attosecond radiation pulse. If the time scale is shorter than the half period of the IR dressing field, the energies of the photoelectrons would be modulated by the Coulomb force exerted by the IR electric field. Thus the photoelectrons' energy distribution in the final state is mainly determined by the phase of the IR field at the instant moment of ionization, and this is called "streaking regime"[5]. Then the temporal properties of the attosecond pulse could be extracted from the time-resolved streaked spectrogram of final photoelectrons. Moreover, the streaking technique could also be utilized to characterize the free electron laser (FEL）pulse which is nominally in the time scale of a few to few tens of femtoseconds, significantly longer than the half period of IR laser. At this circumstance, the IR dressing pulse could be replaced by a terahertz (THz) pulse to maintain the streaking condition[6].

While the above case implicates another situation, where the ionizing pulse e.g. a FEL pulse or a train of isolated attosecond bursts, is much longer than the half period of the IR dressing field. The interplay in-between the ionized photoelectrons absorbing or releasing a photon from the dressing laser, would lead to the coherent interference and intra "sideband" feature with respect to the main peak in the dressed photoelectron spectroscopy. And this represents the so-called "sideband regime"[7, 8]. The sideband peak intensity changes associatively with the time delay in-between the ionizing field and dressing field, and the time-dependent sideband spectrogram could be used to retrieve attosecond pulse train[9]. But this article mainly addresses the optical nature and relevant problems in the "streaking regime". And the streaking technique could be implemented in not only characterization of attosecond pulses but also investigation of the electron dynamic properties in atomic system as well[3, 10, 11].

Generally, the attosecond pulse retrieval techniques can be divided in two sub-categories: i) generalized projection algorithms e.g. frequency resolved optical gating for complete retrieval of attosecond bursts (FROG-CRAB), and ii) frequency phase interferometric methods[12, 13] e.g. phase retrieval by omega oscillation filtering (PROOF), and reconstruction of attosecond beating by interference of two-photon transitions (RABBITT). Mean-while, FROG-CRAB and PROOF belong to the "streaking regime"; but unlike FROG-CRAB, PROOF is restricted in the perturbative regime, where the intensity of the dressing field yields to the slowly varying envelope approximation[13]. On the other hand, RABBIT method satisfies the "sideband regime", which could retrieve the train of attosecond pulses similar as an isolated attosecond pulse[12]. And it needs to point out, the major difference of FROG-CRAB, with respect to PROOF and RABBIT: The algorithm adopts the central energy approximation (CEA) to de-couple the temporal gate function from momentum dependence, very similar to the conventional FROG[14]. But one of the primary consequences due to the CEA scheme is the corresponding bandwidth of the retrieved pulse shouldn't be too broad, otherwise the scheme would introduce excess errors[15].

The current manuscript introduces a dynamic streaking model to simulate the time-resolved spectrogram for ultrashort photoelectron pulses in a dressing field based on the actual electron wave-packet distributions, and develops a brand-new numeric method named "equal value retrieving" (EVR) algorithm to retrieve the attosecond pulse. The model doesn't implement CEA, which would inevitably introduce non-negligible system errors in the simulation and fitting procedures, especially for extremely short pulse e.g. sub-100as. And upon the investigation of the spectrogram for attosecond pulses with different chirps, we revealed and analyzed the correlation in-between the intensity peak distribution of the spectrogram and the chirp of the pulse, to retrieve the detailed chirp profiles along with the pulse lengths simultaneously. Eventually, we demonstrated this brand-new pulse retrieval process allow one to calibrate the temporal properties of sub-100as pulses more accurately.

## 2. Method
### 2.1 Dynamic model for streaking spectrogram

First, let's discuss about the main principle for attosecond streaking. The essential components of an attosecond streak apparatus consists an electron time-of-flight (TOF) spectrometer in a vacuum chamber, along with a gas jet target simulating the photocathode of a conventional device. When a beam of isolated attosecond XUV pulses is focused on the noble gas jet, the bound electrons of the atoms would be ionized and ejected into vacuum by the XUV photons. Thus, the temporal information of the radiation pulses is encrypted into the momentum distribution of the ionized electrons under the streaking potential provided by an infrared dressing field with linear polarization. The kinetic energies (or momenta) of photoelectrons could be measured by the electron TOF spectrometer to record the time-resolved spectra at various delay-times (or phases) in IR dressing field. The photoelectron pulse is considered as a replica of the attosecond pulse, and the electrons' momentum distribution maps the spectrum of the attosecond pulse, according to Einstein's photoelectric effect. Applying the strong field approximation (within non-perturbation regime), the ionized electrons move and behave as classical particles, and their momenta can be calculated by Newton's mechanical laws. If a bound electron of the atom is ionized and released to the IR dressing field at time-delay $t_0$, its final kinetic energy (measured parallelly to the linear polarization of IR field) could be described by,

$$E_p = \frac{q}{m_e} \vec{p_0} \cdot \vec{A}_{IR}(t_0) + q_e^2 \frac{A_{IR}^2(t_0)}{2m_e} + \frac{p_0^2}{2m_e} \tag{1}$$

Where $\vec{p_0}$ is the initial momentum of the freed electron, $\vec{A}_{IR}$ is the vector potential of IR laser field, defined as the time integration of the IR electric field[5],

$$\vec{A}_{IR}(t) = -\int_t^{+\infty} \vec{E}_{IR}(t)dt \tag{2}$$

The final energy $E_p$ could be calculated according to that, for a photoelectron with any initial energy at any arbitrary ionization time $t_0$. Generally, for an attosecond Gaussian pulse with linear chirp, the corresponding spectrum of the pulse can be calculated from its electric field in time domain via short-time Fourier transform (STFT). Fig.1 applied STFT within various timing segments to construct a three-dimensional plot, illustrating the initial time-energy distribution of the ionized electron wave-packet (refer to supplementary materials for more details). The two curves in the horizontal plane present the attosecond pulse's normalized field envelop in time domain (blue), and spectral envelope in frequency domain (red). In the simulation, the photon energy of the attosecond pulse is set to be 200 eV (6.2nm in wavelength), the IR dressing field's wavelength is 800nm (optical cycle of ~2.67as), and Argon is chosen as the gas medium (with ionization potential of 15.76eV).

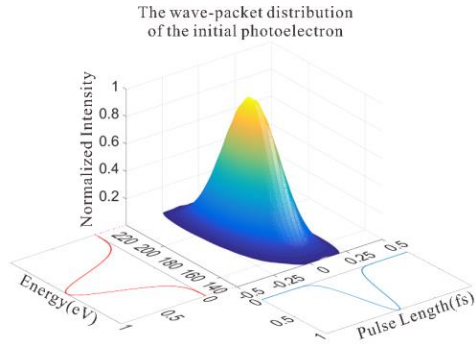

Fig. 1. The 3D feature illustrates the time-energy distribution of the initial photoelectrons, where the time duration of the wave-packet is directly mapping that of the isolate attosecond pulse, and the spectral distribution is calculated through short-time Fourier transform of the attosecond pulse to reconstruct the electrons' wave-packet. The 2D plots sit beside the axes in the horizontal plane are the amplitude envelopes of the wave-packet in time (blue) or energy (red) domains respectively.

Implementing the wave-packet distribution of the initial photoelectrons to Eq.1 for the final kinetic energy, the streaking spectra at various time delays $t_0$ could be calculated. The typical simulated results of the streaking spectroscopy are shown in figure 2. The color graph represents the final kinetic energy distributions for photoelectrons in all initial states, i.e. at different ionization times or with different initial kinetic energies, which were set as the axis of abscissa or ordinate for the color graph respectively. The plot below each color graph shows the IR dressing field and its vector potential i.e. the field's integration in time $\vec{A}_{IR}$, which reflects the distribution of the photoelectrons' final energies is modulated by the vector potential of IR laser field. The wave-packet distribution of the initial electrons spanning a range of kinetic energies, indicates an attosecond pulse ionizing the target atoms at certain delay-time. The wave-packet distribution of the initial photoelectrons could be divided into multiple subdomains with identical colors in the color graph, representing the photoelectrons within those initial states would be modulated to receive the identical final kinetic energy. And the subdomains possessing the same colors are called "region of equivalent interaction". Integrating the intensity of the wave-packet distribution in each of those "regions of equivalent interaction" sequentially, a new wave-packet distribution of photoelectrons in the final state would be generated, and the bar chart on the right side of the color graph in figure 2 is corresponding to the specific photoelectron spectrum at the current time delay. More specifically, Fig.2(a) or (b) is associated with two representative streaked spectra at the maximum or zero crossing of the vector potentials respectively. If this procedure is repeated for each delay successively, a complete time-resolved streaking spectrogram could be obtained (refer to Fig.3). Since this streaking spectrogram is simulated based on the semi-classical mechanism utilizing the electron dynamic evolution in space of time and energy, we named it "dynamic streaking model".

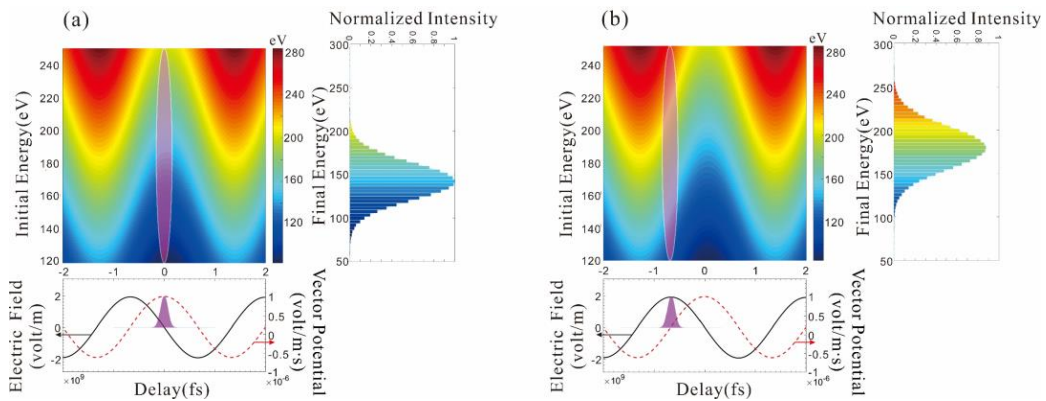

Fig. 2. The schematic of the ultrafast dynamic model. The color figure demonstrates the IR-streaking for the isolate attosecond pulse at various delay-times, where the same color represents the identical final energy (scale bar). The inserted ellipse in diagram (a or b) is the time and energy projection of the wave-packet of the initial photoelectrons and the bar chart on the right side of each shows the normalized photoelectron intensity distribution for the final state, which results from the integration for the wave-packet of the initial photoelectrons within the 'region of equivalent interaction' (see text). The plots below the color graphs illustrate the attosecond pulse streaked by IR electric fields (solid line) along with the vector potentials (dashed line) at two typical time delays, (a) at the maximum of the vector potential; (b) at the zero-crossing of the vector potential.

In Fig.3, the streaking spectrogram by the dynamic model we developed is compared to the traditional FROG-CRAB algorithm, where FROG-CRAB applies CEA while the dynamic model concerns the actual time-energy distribution in the electron wave-packet. And the electron energy bandwidth broadening by FROG model is negligible (20.4eV in FWHM), respected to the non-streaked photoelectron spectrogram (20.1eV in FWHM). While the dynamic streaking model counts the full time-energy distribution of the electrons, resulting an obvious spectral broadening (36.7eV in FWHM) in the spectrogram. Mainly limited by CEA in traditional FROG modeling, the electron energy modulation (by IR dressing field at various delay-times) is dominant, rather than the actual spectral broadening. Especially when the model is adopted to simulate the ultra-short attosecond pulses e.g. sub-100as, it would introduce excess errors, thus CEA would be no longer applicable.

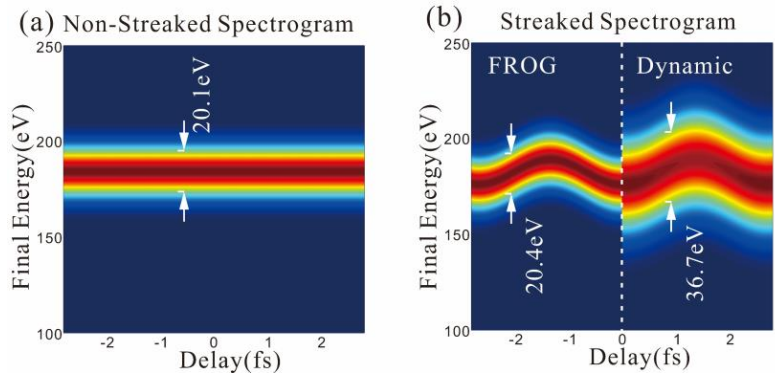

Fig. 3. presents the streaking spectrogram calculated by different models. (a) is the non-streaking spectrogram, whose spectral width in FWHM is about 20.1eV; (b) shows the streaking spectrum calculated by FROG model (left side) and by dynamic model (right side). Their spectral width (FWHM) of the IR streaking spectrograms are demonstrated, where the electron energy spectral broadening for FROG model is negligible while the broadening effect for dynamic model is substantial.

*2.2 Equal-value Retrieval Method*

The dynamic model is used to generate the streaking spectrogram, which concerns the actual time-energy distribution of the initial photoelectrons' wave-packet in IR dressing field. The simulated streaking spectrograms for attosecond pulses with identical pulse length of 90as but different chirps are presented in Figure 4, displaying very different features, which would be analyzed further shortly.

Fig.4 (a-c) show the photoelectrons' final energy distributions for pulses with various chirps and their dynamic streaking processes; (d-e) are the corresponding streaking spectrogram for each of them. More specifically, the color graph in Fig.4(a) and the spectral plot right below it (Fig.4(c)) are the final energy distribution profile and the streaking spectrogram respectively, for a pulse with positive chirp. Where the positions of the local maximum (darker region) and minimum (lighter region) of the spectral intensity are highlighted in the spectrogram (Fig.4(c)), which are linked to the corresponding initial photoelectrons' wave-packets (tilted transparent ellipse) in the color graph (Fig.4(a)) associatively, to demonstrate how the dynamic streaking modulates the electrons' initial energy distribution to generate the final one. And it is not difficult to observe that the local maximum of the streaking spectral intensity would appear at the time scale when the final energy distribution is in a same phase as that of the photoelectrons' initial wave-packet. Since the electrons in initial state with positive chirp, confined in such a time-energy domain, would contribute to the dynamic streaking effect constructively. Thus, we define it as "region of equivalent interaction in positive chirp" (white dot-line), corresponding to the spectral local maximum for positive chirp pulses. On the other hand, if positive chirp pulses in initial state are confined in a domain having a counter phase, defined as "region of equivalent interaction in negative chirp" (blue dot-line), the spectral local minimum would appear. Analogously for an attosecond pulse with negative chirp, the location of local maximum or minimum would exchange to vice versa, i.e. the maximum appears at "region of equivalent interaction in negative chirp" while the minimum at "region of equivalent interaction in positive chirp" (refer to Fig.4(b, e)). Furthermore, for a chirp-free pulse in transform-limit, the pattern of spectral intensity distribution is almost uniform for all domains (or delay times), only oscillating coordinately with the optical cycle of the dressing field (Fig.4(c, f)).

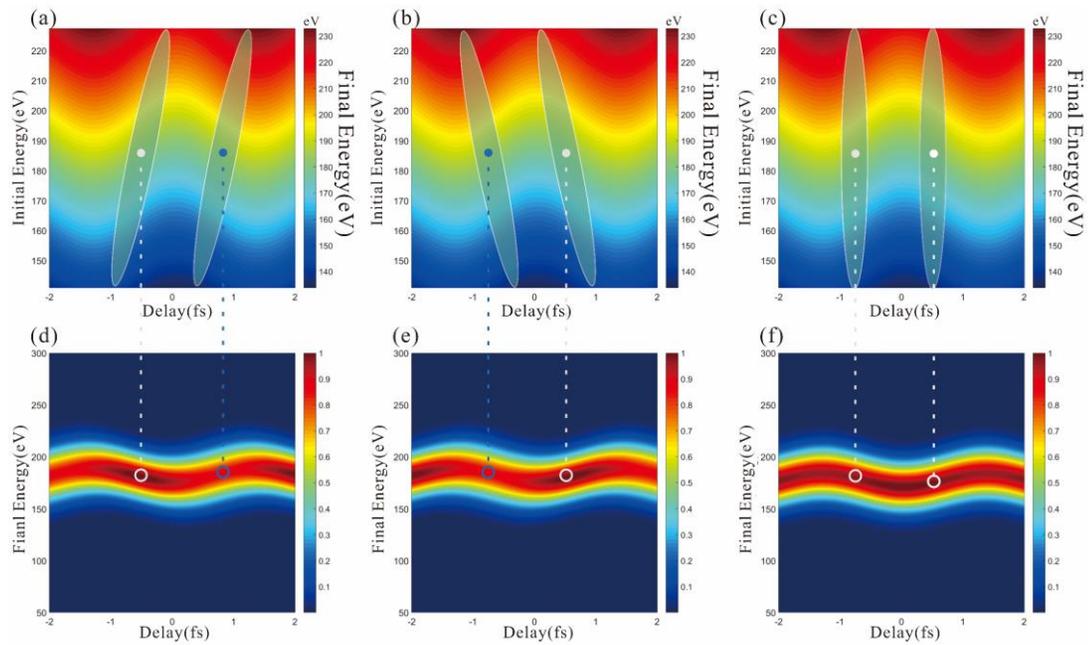

Fig. 4 demonstrates how the dynamic model is implemented to obtain the streaking spectrogram for attosecond pulses with different chirps, the color graphs in the upper row present the schematic of the dynamic model (refer to Fig.2), and the lower row are the corresponding streaking spectrogram for each case (refer to Fig.3). Among them, (a)(d) are the photoelectrons' final-energy distribution profile and the streaked spectrogram for an attosecond pulse with positive chirp respectively; (b)(e) illustrate the same process for an attosecond pulse with negative chirp; (c)(f) the same for chirp-free attosecond pulse.

Therefore, according to this apparently characteristic feature, a rough comparison of the streaking spectrograms would be able to tell the chirp properties of attosecond pulse qualitatively. As demonstrated in Fig.4, if the ionization takes place within a "region of equivalent interaction" whose chirp has the same sign with that of attosecond pulse, the streaking spectrogram would take a local maximum, while take a local minimum for the vice versa. Based on that, we came up a fresh idea to analyze the related parameters of the streaked spectrogram and use them to reconstruct the attosecond pulses. Primarily for a given attosecond pulse with known pulse length and chirp, the streaked spectrogram could be simulated via the dynamical model when scanning the delay times in IR dressing field, then three essential parameters could be retrieved from the spectrogram: (1) the ratio of the local minimum to local maximum of the spectral intensity $I_{min}/I_{max}$; (2) the delay time of the local maximum with respect to the time zero $\tau_{max}$; (3) the spectral width (FWHM) of the spectrogram at the delay for local maximum. Thus, various pulse lengths and chirps could be used to simulate various specific spectrograms, from which the three aforementioned parameters could be retrieved to construct their own distribution profiles; and in each of them, the pulse length and chirp are the variables. Again, from the streaked spectrogram of a test attosecond pulse, the expectation values of the three parameters could be retrieved independently, forming the specific planes of "equal value" to intersect the corresponding distribution profile surfaces respectively. Thus, the intersection lines would be generated and projected to the plane of "pulse length – chirp". And these would represent the primary constrains for reconstruction of the pulse duration and chirp of ultrafast attosecond pulses. More concrete discussions would be given in next.

Figure 5 presents the distribution profiles for the three essential parameters, retrieved from a series of streaked spectrograms, spanning certain domain in the plane of "pulse length - chirp". Especially, the plot on the right side of Fig.5(a) shows the curves of "$I_{min}/I_{max}$ vs. chirp" at three different pulse lengths: 40, 70 and 100as, related to the crossing lines by the vertical cutting planes at the specific pulse lengths to the distribution surface. And we find: i) As the chirp value approaches to zero, the ratio of $I_{min}/I_{max}$ is getting very close to but a bit smaller than 1. This is due to "region of equivalent interaction" is more concentrated at the zero crossing of vector potential, causing the slight spectral intensity fluctuation along the time axis even for an ideal chirp free pulse. ii) $I_{min}/I_{max}$ declines quadratically with chirp, while the quadratic curves would become more flat for shorter pulse length, indicating it would be more difficult to characterize the chirp for shorter pulse, e.g. sub-100as. The distribution profiles for the other two quantities (time delay and spectral width at the local maximum of the spectrogram) are displayed in Figure 5 (b, c).

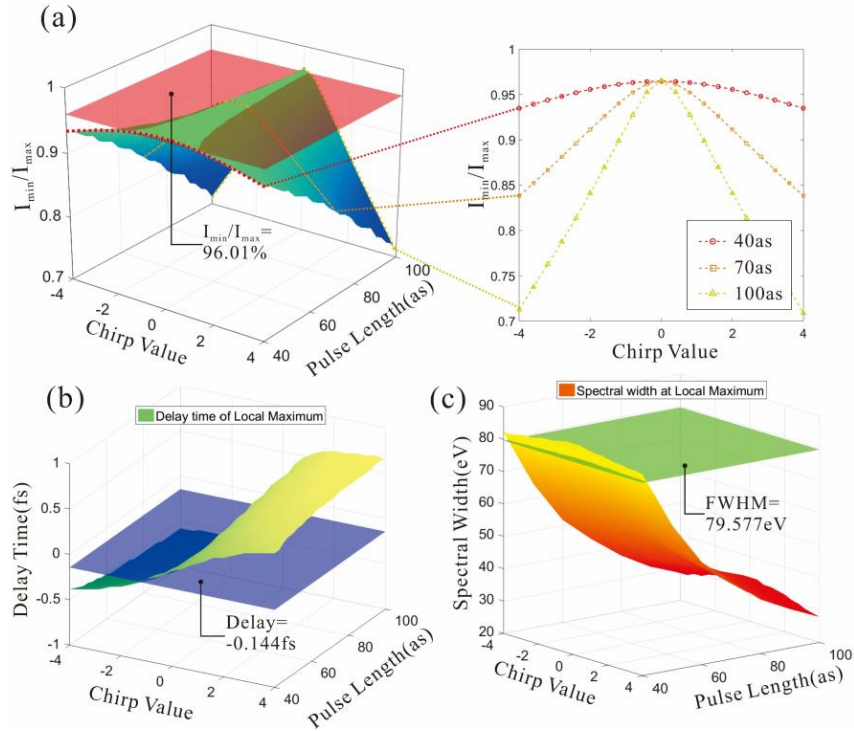

Fig. 5. The featured profiles for the essential parameters extracted from IR-streaking spectrogram (Fig. 4) for attosecond pulses with various pulse-lengths and chirps. (a) is the distribution profile for the ratio of the local minimum to local maximum of the streaking spectral intensity, $I_{min}/I_{max}$; three curves on the right side are the intersection lines at three attosecond pulse lengths, 40as, 70as and 100as. (b) is the distribution profiles for the time delay of the local maximum of the streaking spectrogram with respect to the time zero. (c) is the distribution surface for the spectral bandwidth (in FWHM) at the local maximum of the streaking spectrogram. Each 'plane of equal-value' crossing the three distribution profiles, using the specific parameters retrieved from the simulated streaking spectrogram for a test attosecond pulse via ultrafast dynamic model (refer to Fig. 4).

Our method to reconstruct the attosecond pulse is based on the dynamic model and refer to the essential distribution profiles crossing with the specific planes of equal values, therefore we named it "equal value retrieval" (EVR) method. Upon projection of all these intersection lines on the plane of "pulse length - chirp", the temporal profiles of attosecond pulses could be identified. There are four major steps for this new method. 1) Construction of the distribution profiles for the three parameters mentioned before, which could be acquired from the streaking spectrograms via dynamical model, by implementing series of pulse lengths and chirps. 2) From the measured streaking spectrogram (e.g. from a test attosecond pulse), the values of the three parameters could be calibrated, forming the planes of equal-values in each distribution profiles. 3) Each plane of "equal-value" cuts through the specific distribution profile to generate "curve of equal-value". 4) "Curves of equal-values" for the three characterization parameters are projected to the plane of "pulse length-chirp" and intersect each other, and the intersection points could be used to identify the retrieved attosecond pulse's temporal profiles. As demonstrated in Fig. 5, "curves of equal-values" in $I_{min}/I_{max}$ (red) and the spectral width at the local maximum (green) were mainly used for identification of the intersection point, since the absolute value of "delay" $\tau_{max}$ is rather difficult to calibrate precisely in sub-fs time scale. However, the curve of equal value - delay (blue) is also very useful for determining the sign of the chirp, either positive or negative. This gives only one unique set of coordinates for the intersection point in Fig. 6, corresponding to the retrieved attosecond pulse.

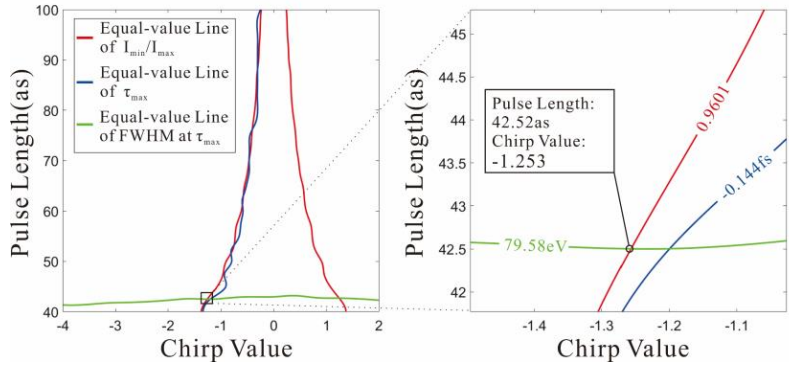

Fig. 6 The retrieval algorithm "curve of equal-value" for attosecond pulse calibration. Three solid lines in the figure are the equal-value line corresponding to the intensity ratio ($I_{min}/I_{max}$, red), the phase delay at the local spectral maximum $\tau_{max}$ (blue), and FWHM energy width at that delay (green), which are extracted from various distribution profiles crossed by contour surfaces at certain equal values in Fig.5. The right panel is the 'zoom-in' feature for the intersection region of the left panel, represented by a small rectangle.

### 3. Discussion and Conclusion

Fig.4(d-e) mainly present the typical streaked spectrogram for a 90-attosecond pulse with positive, negative or free chirps, while the chirp related feature would become less distinct for shorter pulses. As shown in Fig.5(a), the quadratic curves of "$I_{min}/I_{max}$ vs. chirp" are gradually flattened out for shorter pulse lengths, indicating calibration of extremely short pulses e.g. sub-50as is much more challenging. Thus, in Fig.5 and 6, we implemented a test sub-50 as pulse and demonstrated the performance of the dynamic simulation model and EVR pulse reconstruction scheme. And the test XUV attosecond pulse signal in Gaussian profile carrying a linear chirp is given by:

$$E_{XUV}(t) = e^{-2\ln 2 \cdot t^2 (1+ib)/[(1+b^2)\tau^2]} \qquad (3)$$

Where the pulse duration $\tau$ in FWHM is 43as and the unitless chirp parameter b=-1.244, identical to the chirp value in conventional format via $c = -2\ln 2 \cdot b/[(1+b^2)\tau^2] = 3.661 \times 10^{-4} as^{-2}$. The retrieved results for this attosecond pulse by the traditional FROG-CRAB model or dynamical model in EVR scheme are present in Fig.7, where Fig.7(a) shows the field amplitude envelope for each of them, Fig.7(b) presents their phases, and the retrieved results are compared with the original test signal in same plots. Obviously at this circumstance, the FROG-CRAB results deviate from the original signals significantly: the pulse length of the retrieved amplitude envelope is only half of the actual value, the retrieved chirp is negative for an original positive chirped pulse. In contrast, the EVR method based on the dynamic streaking model, displays excellent retrieval results, with a pulse duration for the amplitude envelope of 45.52 as and a chirp value of -1.253, which are very close to the original values. Since FROG-CRAB adopts CEA for the initial photoelectron wave-packet, which would introduce considerable system errors if the pulse duration goes down shorter and shorter to sub-50as, leading to large discrepancy in-between the retrieval results and the original ones. While in the dynamical model, the authentic time-energy distribution for the initial photoelectron wave-packet are concerned in the simulation, and the EVR method could well discriminate the actual constraints for those essential parameters' distribution profiles, therefore the retrieved results are much more accurate, compared to the conventional techniques applying CEA.

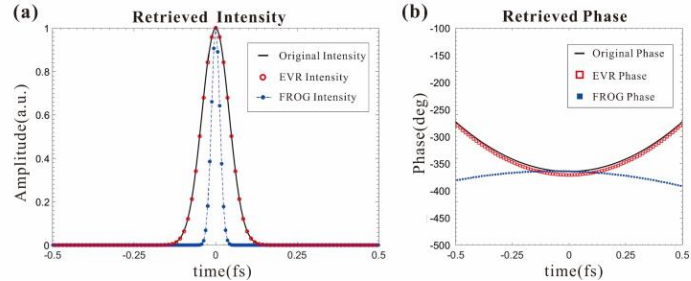

Fig. 7 (a) The reconstructed attosecond pulse's field amplitude envelopes by EVR (open circle) or FROG-CRAB (filled square) models are compared with the original signal; (b) The reconstructed phases by these two models are compared to the original signal.

With the advancement of attosecond science and technology, it enables to investigate the frontier sciences and ultrafast processes within the time scale of few attoseconds[16]. Shorter attosecond pulses and higher photon energies are in great demand for those experimental approaches at the cutting-edge. In this work, we developed a dynamic streaking model to describe the photoelectrons' time-energy evolution within the dressing field and to simulate streaking spectrogram. According to thoroughly systematical analysis of the spectrograms, we selected a few parameters as the constraints for pulse reconstruction; and based on that, a novel scheme of equal-value-retrieval method (EVR) is established. Since our model avoids the central energy approximation (CEA) and identifies the correlations in-between the pulse profiles (pulse length and chirp) and the features of spectrogram, it could characterize the attosecond pulse profiles much more precisely respected to the traditional methods, especially well applicable for sub-50as ultrashort pulses. While this method requires more time and resource for computation due to the complexity in modelling, simulation, and searching algorithm, but remarkably it brings up an original, creative and reliable approach to calibrate sub-50as pulses!

## Funding

This work was supported by the National Natural Science Foundation of China (Grant: 11475249) and Youth 1000-Talent Program in the China (Grant: Y326021061).

## Acknowledgments

The authors thank for the staff and facility support from the Department of Free Electron Laser Science and Technology, Shanghai Institute of Applied Physics, Chinese Academy of Sciences.

**Supplementary Materials**

*Short-time Fourier transform*

The short-time Fourier transform (STFT), also called short-term Fourier transform, is a Fourier transform process applied to the time-domain signal. But unlike the regular Fourier transform, the time scale for the time-domain signal in STFT is divided into multiple shorter segments in equal length, and Fourier transform is then processed individually and sequentially within each of these segments. In doing so, the transformed result i.e. the spectrum for each specific segment could be plotted against time, forming a 3D distribution profile as a function of variables: time and energy,

$$STFT\{x(t)\}(t',\omega) = \int_{-\infty}^{\infty} x(t)w(t-t')e^{-i\omega t}dt .$$

Where w(t) is the function of temporal gate, usually in Hann or Gaussian profile; x(t) is the time-domain signal to take the Fourier transformation; the spanning time range of t' is normally much longer than the timing step 'dt' during the calculation, thus it could be treated as a "slow" or "quasi-static" variable.

As discussed in the main text, the temporal distribution profile of initial photoelectrons mapping the field envelope of an ionized XUV attosecond Gaussian pulse with linear chirp, could be expressed as,

$$E_{XUV}(t) = e^{-2\ln 2 \cdot t^2 (1+ib)/[(1+b^2)\tau^2]} .$$

Through STFT, the wave-packet of the initial photoelectrons is constructed, and the chirp feature i.e. "frequency change vs. time" could be clearly observed (refer to Visualization 1). The scheme is completely different with the traditional Fourier transform to the signal spanning the whole time scale, and the concrete procedures could be found in Visualization 2.

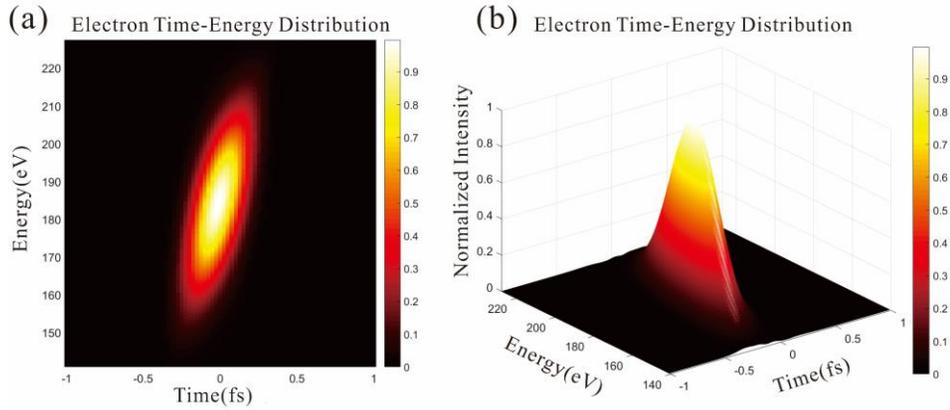

Visualization 1. The 3D wave-packet for initial photoelectrons' time-energy distribution, generated by STFT from the field envelope of a 90as pulse with positive chirp. Fig(a) presents the time-energy distribution plot, where the obvious positive chirp feature mapping from an identical ionizing pulse could be observed; Fig(b) is the 3D diagram of Fig(a).

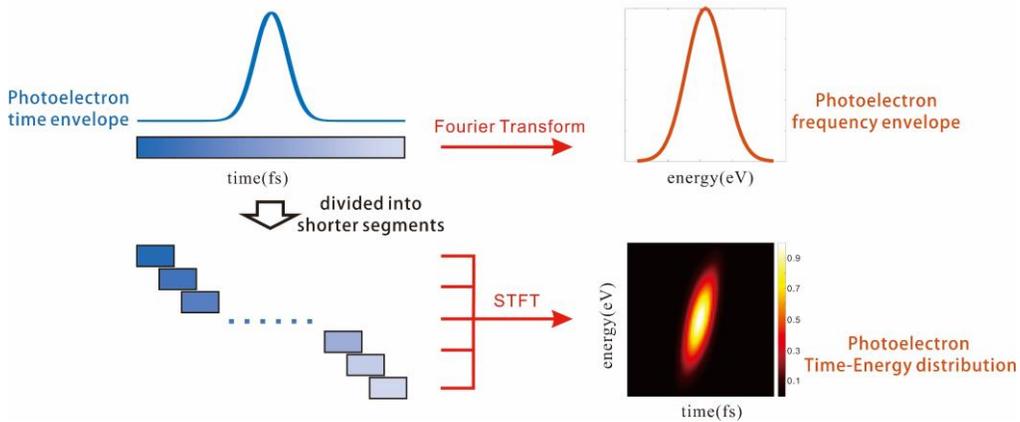

Visualization 2. The schematic scheme of STFT to create the wave-packet for the initial photoelectrons stimulated by an attosecond pulse with positive chirp. In the process, the field envelope of the photoelectrons is treated as the time signal, and the time scale is divided into multiple shorter segments in equal length via timing gate function. The time signal within each segment contains local information of field magnitude and phase, thus STFT could reflect the frequency within a local time and its dynamic evolution over time. And the 3D time-energy distribution profile shows the electrons' wave-packet maintains a positive chirp induced by an ionizing pulse.